# Classification of Hand Gestures from Wearable IMUs using Deep Neural Network

Karush Suri*, Rinki Gupta

Electronics and Communication Engineering Department
Amity University Noida, Uttar Pradesh-201313, India
karushsuri@gmail.com, rgupta3@amity.edu

*Abstract*— IMUs are gaining significant importance in the field of hand gesture analysis, trajectory detection and kinematic functional study. An Inertial Measurement Unit (IMU) consists of tri-axial accelerometers and gyroscopes which can together be used for formation analysis. The paper presents a novel classification approach using a Deep Neural Network (DNN) for classifying hand gestures obtained from wearable IMU sensors. An optimization objective is set for the classifier in order to reduce correlation between the activities and fit the signal-set with best performance parameters. Training of the network is carried out by feed-forward computation of the input features followed by the back-propagation of errors. The predicted outputs are analyzed in the form of classification accuracies which are then compared to the conventional classification schemes of SVM and *k*NN. A 3-5% improvement in accuracies is observed in the case of DNN classification. Results are presented for the recorded accelerometer and gyroscope signals and the considered classification schemes.

I. INTRODUCTION

Distinguishing the various hand gestures effectively plays an essential role in applications ranging from gesture recognition to kinematic study [1]. A greater scope is observed in healthcare development and diagnostic learning problems wherein an accurate estimation of the activity of the muscle is required. The analysis may be used for treating muscle abnormalities and restricted motor functioning [2]. IMU sensors capable of recording the muscle activity during the motion of the hand are utilized for this purpose. These sensors record signals assessing the degree of acceleration and rotation during the interval in which the activity is performed. Signals may then be processed and analyzed for the study of hand gestures.

Analyzing signals containing information about acceleration and rotation with respect to time play a key role in activity recognition due to their three dimensional nature. These signals are obtained by the virtue of muscle action and can be analyzed individually or as a combination of multiple sensor inputs. Multiple sensor systems maybe used together which can be combined with a classifier for recognition [3]. Multiple classification and activity detection algorithms may be used for analyzing wearable sensor signals. Gaussian models and Deep Belief Networks (DBNs) are the most commonly used approaches for improved learning in literature. Due to multiple layers introduced in the network, DBNs improve learning by means of increased computations. When compared to commonly used classifiers such as Support Vector Machines (SVM), Logistic Regression (LR) and *k*-Nearest Neighbors (*k*NN), DBNs outperform the other models in terms of complexity and accuracy [4]. Using a simple Artificial Neural Network (ANN) provides efficient classification. However, due to the usage of only a single layer in ANN, the network may still be considered a shallow classifier [5]. Implementation of ANN for recognition of basic activities is suitable in literature due to the fact that the complexity of input is low. The network may function effectively by means of dominant features which play a significant role in reducing correlation. This provides a basis for correctness of the network [6-7]. On modeling the network as a multiple hidden layer network, commonly known as a Deep Neural Network (DNN), improvement in the learning process may be obtained. Addition of layers in the DNN provides a better fit for the input data-points by implementation of the activation function at each node [8]. Effect of variation in the number of hidden layers of the DNN can be also be observed in the construction of the model. With increasing depth of network, the layers detect more intricate aspects of the activities [9]. Combining DNNs with unsupervised approaches may depict suitability in literature. The feature transforms considered should consist of dominant features and discriminative techniques capable of yielding an accurate model [10].

In order to carry out classification in a more effective way, use of an optimized classifier algorithm is essential. Starting with a basic structure, the structure of the classifier may be improved by achieving a pre-defined optimization objective. The objective may be achieved with the help of a control parameter capable of governing the learning ability of the algorithm. Once the construction of the design is complete, the algorithm may be used to classify the various input IMU signals into their respective activity classes. Validity of the proposed classifier algorithm may then be assessed by comparing its performance with the conventional ones.

In this paper, an improved classification approach is proposed by utilizing a novel DNN for classification of six gestures of hand corresponding to signs in the Indian Sign Language using signals recorded from three IMUs each consisting of tri-axial accelerometer and gyroscope placed on the forearm. Classification of gestures is carried out by the proposed approach and validated by comparing the results to that of conventional classifiers such as SVM and *k*NN. The remaining paper is organized as follows. Section II describes the features considered for classification and contains a brief description of the feature reduction and classification techniques used in this work. The proposed DNN classification algorithm is presented in Sec. III. Results depicting the correctness of proposed grouping of the activities and the classification of six hand gestures using IMU signals are given in Sec. IV. Section V provides the conclusion of the paper.

## II. CLASSIFICATION USING CONVENTIONAL ALGORITHMS

### A. Data Corpus

The signals from accelerometer and gyroscope have been measured by placing 3 wireless Delsys IM sensors on the dominant hand of the subject as shown in Fig. 1. The objective is to utilize the signals to classify six gestures performed by the right hand. The signals have been collected from 4 healthy subjects in the age group of 22-30 years, all right-handed females. The six gestures that are performed are signs for Win, Loose, Key, Bold, Confident and Sorry from the Indian Sign Language published by the Faculty of Disability Management and Special Education (FDMSE) of Ramakrishna Mission Vivekananda University (RKMVU), Coimbatore, India. An audio stimulus of 3 seconds is played to the subject to indicate when the gesture is to be performed, with 5 seconds of rest in between each gesture. Each subject performed 20 repetitions for each gesture and the signals are recorded in continuation as one recording. The accelerometer and gyroscope signals are obtained with a sampling rate of 148.15 Hz and a bit depth of 16 bits. A delay of 0.5 seconds has also been taken into account generated by the recording apparatus and compensated as hardware delay.

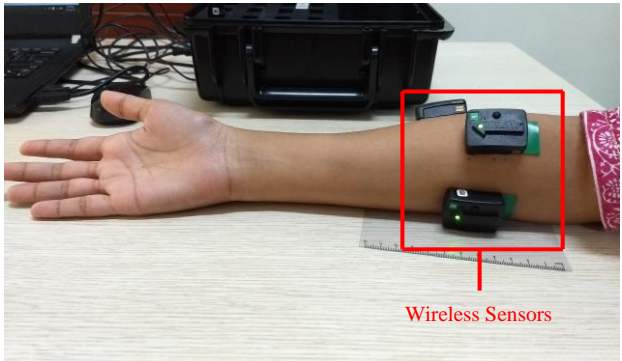

Fig. 1. Accelerometer and gyroscope sensors placed on the forearm

### B. Features for analysis of IMU signals

The most prominently used time-domain features are considered here [11]. These include:

*1) Mean*: Mean is defined as the average value of the number of instances over a specific period. Mean of a random variable $x$ having $n$ values can be expressed as

$$\text{Mean} = \frac{1}{n}\sum_{i=1}^{n} x_i. \quad (1)$$

*2) Mean absolute value (MAV)*: Mean absolute value (MAV) is the absolute value of the mean of a segment of data. MAV of a random variable $x$ having $n$ values can be expressed mathematically as

$$\text{MAV} = \left|\frac{1}{n}\sum_{i=1}^{n} x_i\right|. \quad (2)$$

*3) Standard Deviation*: Standards deviation is defined as the square root of the variance

$$\sigma = \sqrt{\frac{1}{n-1}\sum_{i=0}^{n}(x_i - \mu)^2}, \quad (3)$$

where $\mu$ is the mean of the data.

*4) Root Mean Square (RMS)*: The RMS value of a data segment is the root of the mean of squares of the data values. The RMS of a random variable $x$ having $n$ values can be expressed mathematically as

$$\text{RMS} = \sqrt{\frac{1}{n}\sum_{i=0}^{n} x_i^2}. \quad (4)$$

*5) Variance*: Variance of a random variable $x$ is defined as the square of its standard deviation value. It can be mathematically expressed as

$$var(x) = \frac{1}{n-1}\sum_{i=0}^{n}(x_i - \mu)^2. \quad (5)$$

*6) Waveform Length (WL)*: It is a cumulative variation of the sEMG that depicts the degree of variation about the signal, given as

$$\text{WL} = \sum_{k=1}^{n-1}|x_{k+1} - x_k|. \quad (6)$$

*7) Auto-Regressive Coefficients:* These are the constant coefficients ($a_r$) of the auto-regressive (AR) model $\hat{x}$ of the sampled instants of the sEMG signal, given as

$$\hat{x}_n = \sum_{k=1}^{p} a_k x_{n-k} + \varepsilon, \quad (7)$$

where $p$ is the order of the AR model and $\varepsilon$ is the prediction error.

*8) Skewness:* The skewness of any signal segment with variable value $x$, having mean $\mu$ and standard deviation $\sigma$ is given as

$$\gamma = \frac{E\{x - \mu\}^3}{\sigma^3}, \quad (8)$$

where $E$ is the expectation operator.

*9) Mobility*: Mobility is defined as the ratio of the variance of the first derivative of the segment $x$ to the variance of the segment, given as

$$\text{M} = \frac{\nabla(var(x))}{var(x)}. \quad (9)$$

*10) Kurtosis:* The kurtosis of any signal segment with variable value $x$, having mean $\mu$ and standard deviation $\sigma$ is given as

$$\kappa = \frac{E\{x-\mu\}^4}{\sigma^4}. \quad (10)$$

*C. Feature reduction and classification*

Dimensionality reduction leads to an improvement in the classification performance due to reduction in correlation in the feature set. Using the features mentioned above in the Principle Component Analysis (PCA) based feature selection process, the number of features is experimentally reduced to a minimum while maintaining the classification accuracies as high as possible. The PCA algorithm projects the *d*-dimensionality data onto *l* eigen-vectors corresponding to their covariance matrix, thus leading to a linear transformation of the original space [12]. The PCA algorithm computes the principle arguments in two mutually orthogonal directions by simplifying the multivariate data. Reduction in the dimensionality serves as an important step in order to predict the values for multiple observations with similar characteristic traits.

The SVM maps the inputs implicitly to its corresponding outputs by making use of the most optimized hyper-plane in the high- dimensional feature space [13]. Non-linear classification in SVM makes use of the kernel trick where a kernel, which is a generalized form of the function is used. This enables the function to map the entities in a high dimensional feature space by successive computation of the dot product. However, in case of six independent activities the method involves the use of masking the residues. Each activity is considered as a separate entity and then distinguished from the remaining counterparts. The replicates contained in the test data are predicted on the basis of which the accuracy of this classification problem is assessed.

The *k*NN algorithm, on the other hand, is a single instance based algorithm wherein local values of the function are computed. Corresponding to each data point in the feature space, neighboring data points are assessed for grouping [14]. If any '*k*' points are classified in the same class, then the concerned data point has a tendency of belonging to the same class. Weights to the contributions of neighbors are assigned corresponding to their distance '*D*' from the data point. KNN classification does not require an explicit training mechanism as the neighbors are taken from a grouping of objects whose class is known [15-16]. Thus, each activity is classified based on the majority grouping of each data point in the signal. Choosing an optimal of '*k*' is of significant importance as large values of '*k*' make the boundaries between the classes less distinct.

On analyzing the results for SVM and *k*NN (discussed in Section IV), further scope of improvement lies in classifying the activities by means of increased training. In the case of SVM and *k*NN, manual control of the training process cannot be monitored. Therefore, a concise idea regarding the amount of training to be provided to the classifier is not achieved. By monitoring the amount of training iterations required and depth of the algorithm, a better fit to the signal-set may be provided. This would help in avoiding confusion given a set a features for classification. Here, grouping is proposed to be carried out on the basis of a deep algorithm whose training iterations may be manually controlled and monitored for improved performance. Construction of such an algorithm requires a pre-requisitely set optimization objective which may be achieved as explained in the following session.

III. PROPOSED DNN HAND ACTIVITY CLASSIFICATION

In this work, classification of six hand activities are considered. As depicted in Fig. 2, classification is carried by making use of the proposed DNN approach and then compared to conventional algorithms such as SVM and KNN. The proposed approach makes use of a DNN which, unlike an ANN, consists of multiple hidden layers capable of learning the intricate aspects of data. Learning of the network takes place in two phases, forward propagation followed by back-propagation.

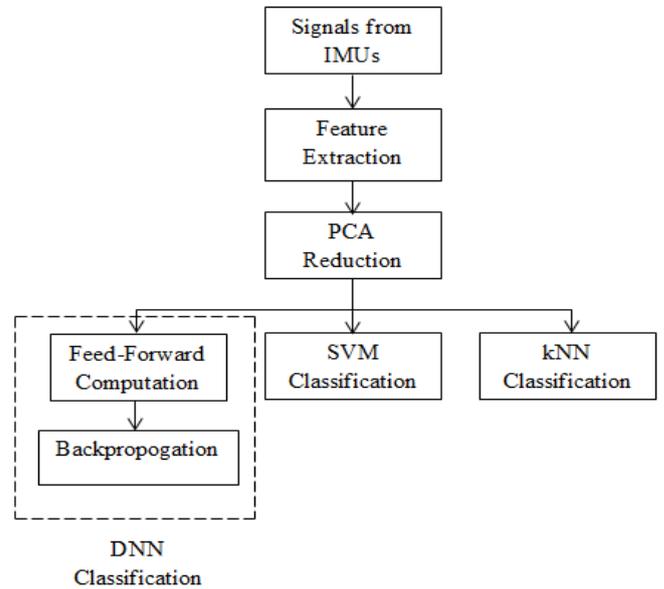

Fig. 2. Proposed DNN approach Vs. SVM and KNN classification

During the forward propagation phase, DNN takes in the input as features of hand activity signals and carries out computations at each neuron of each layer by means of an activation function (sigmoid function in this case). These computations lead to the minimization of the cost function, which is defined as the optimization objective of the network. The activation function and cost function can be mathematically expressed as

$$g(x^{(i)}) = \frac{1}{1+e^{-x^{(i)}}}. \quad (11)$$

$$J(\Theta) = -\frac{1}{m}\left[\sum_{i=1}^{m}\sum_{k=1}^{K} y_k^{(i)} \log(h_\Theta(x^{(i)}))_k + (1-y_k^{(i)})(1-\log(h_\Theta(x^{(i)})))_k\right] \quad (12)$$

Here, '*m*' is the total number of signals to be classified, '*K*' is the total number of classes, '$h_\Theta$' is the hypothesis yielding the prediction '$y^{(i)}$' corresponding to input '$x^{(i)}$'. In the case of multi-class classification, the concerned class is represented as '1' by the hypothesis and the remaining classes as '0'. The hypothesis predicts the values by means of weights 'Θ'. The weights are multiplied by the computed values at each neuron and pass the resultant value to the next layer. The output produced by each layer 'L' may be represented as '$a^{(L)}$'. Over each iteration of feed-forward propagation, the value of the cost function decreases by virtue of optimized values of 'Θ'. Once, ample number of iterations are carried the cost function reaches its minimum value or the global minima. The output layer then produces the final prediction corresponding to the input.

Fig. 3(a) presents the operation of constructed DNN in feed-forward phase along with the design specifications. The input layer consists of 10 neurons and each hidden layer consists of 15 neurons. Number of neurons in the hidden layer are kept approximately half of the number of neurons in the input layer in order to avoid computational complexity. Inputs to the input layer are 10 features ranging from '$F_1$' to '$F_{10}$' obtained from PCA feature reduction. The predicted output is presented at the output layer consisting of only a single neuron.

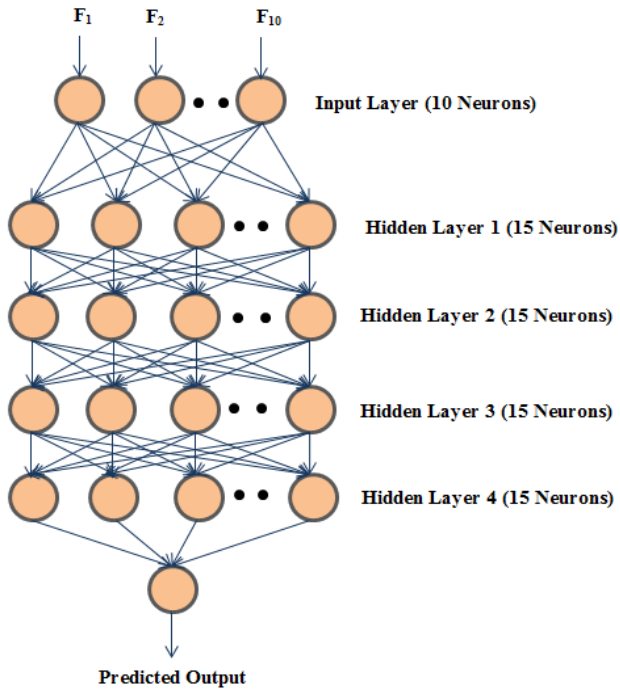

(c) Feed-forward operation of DNN along with design specifications

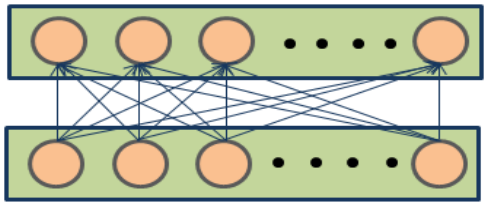

(d) Back-propagation between two consecutive layers

Fig. 3. Design and representation of the two phases of DNN operation for hand activity classification

In the case of back-propagation phase as depicted in Fig. 3(b), the network operates in reverse direction in order to compute the errors '$\delta^{(j)}$' at each neuron '$j$'. Computation of errors begins from the last hidden layer by simply calculating the difference between the output of the layer given as '$a_j^{(5)}$' and the predicted output '$y_j$'. For rest of the layers, the errors are propagated towards the input layer by multiplication of the weights 'Θ'. No errors are produced at the input layer neurons due to the absence of computation. Mathematical expressions for errors at neurons of last hidden layer and at neurons of other layers respectively are represented as

$$\delta_j^{(5)} = a_j^{(5)} - y_j. \qquad (13)$$

$$\delta_j^{(L)} = a_j^{(L)} * (1 - a_j^{(L)}). \qquad (14)$$

For each iteration in DNN, the input undergoes feed-forward propagation and back-propagation. The process is terminated at the achievement of the global minima by the cost function, i.e. when the network is optimized. Predicted values obtained corresponding to each signal '$m$' are used to compute the classification accuracy. Results for DNN classification are compared to the results for SVM and kNN classification and discussed in the next section.

## IV. RESULTS AND DISCUSSION

In order to assess the suitability of the designed DNN, a performance analysis of the network is carried out. The performance of the network under optimized parameters is then compared to the performance of SVM and *k*NN. The accelerometer and gyroscope signals for one repetition of the sign 'sorry' by subject 1 is depicted in Fig. 4(a) and Fig. 4(b), respectively. These are time varying signals and are used as inputs for feature extraction. Each signal consists of its three-dimensional coordinates in the cartesian plane. The extent of acceleration and rotation is differently measured w.r.t. each of these coordinates.

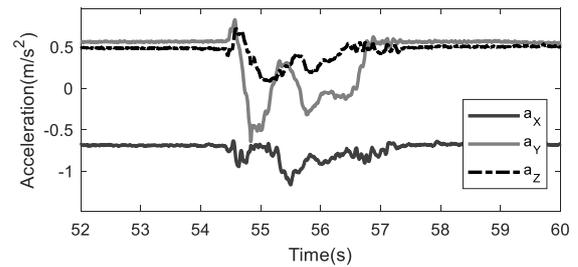

(a) Accelerometer signals for X, Y and Z coordinates

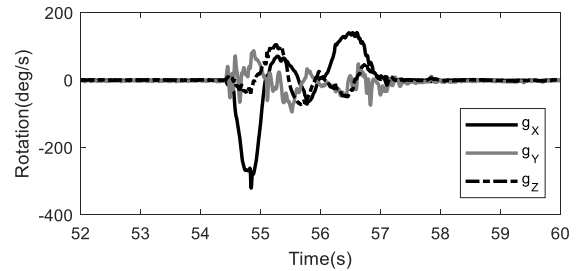

(b) Gyroscope signals for X, Y and Z coordinates

Fig. 4. Input signals for the sign Sorry

Monitoring the performance of DNN is carried out by observing the variation of classification accuracy with the number of iterations for varying number of features as inputs as depicted in Fig. 5. Number of iterations play a significant role in optimization of the algorithm and is thus, essential for assessment. Number of features, on the other hand is used as a control parameter as it directly relates to the input data. As shown in Fig. 5, value of classification accuracy increases with increase in the number of iterations. This validates the descent of cost function towards global minima. On reaching a certain number of iterations, the variation decreases. This indicates the minimized value of the function and signifies that the classifier has been optimized. At 150 iterations, the variation of plot begins to settle which represents the optimal number of iterations. For varying number of features, it is evident that with increase in the number of features the overall values of accuracies on the plot also increase. However, on reaching a certain number of features, say 10, the best accuracies are obtained in fewer number of iterations. This signifies the optimized input for the classifier.

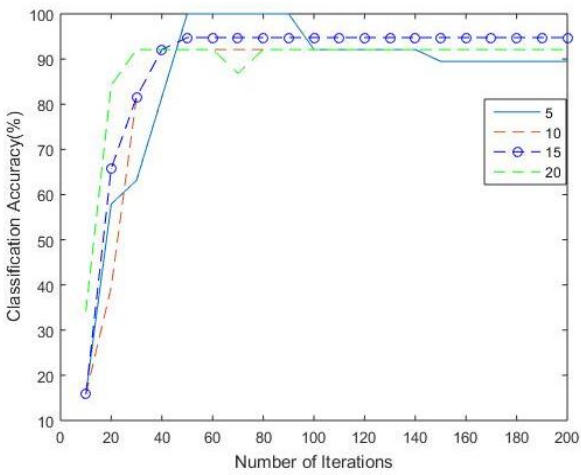

(a) Optimization of classification accuracy for accelerometer input

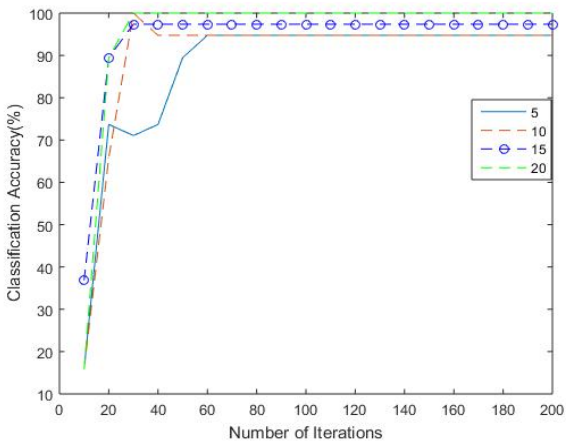

(b) Optimization of classification accuracy for gyroscope input

Fig. 5. Performance analysis of proposed DNN

Once the number of features for DNN is optimized, its classification may be compared to that of conventional algorithms. Comparison is carried on the basis of classification accuracy for each subject for both IMU inputs. As depicted in Fig. 6, the proposed DNN approach performs better in comparison to SVM and $k$NN for both the considered IMU inputs. Peak value of classification accuracy for DNN in the case of accelerometer is observed to be 97.368% whereas for $k$NN the peak value is 92.982%. Peak value for SVM is observed to be 94.737%. In the case of gyroscope, the peak value of classification accuracy is observed to be 99.123% and for $k$NN the peak value is 99.123%. For SVM a peak value of 96.491% is observed.

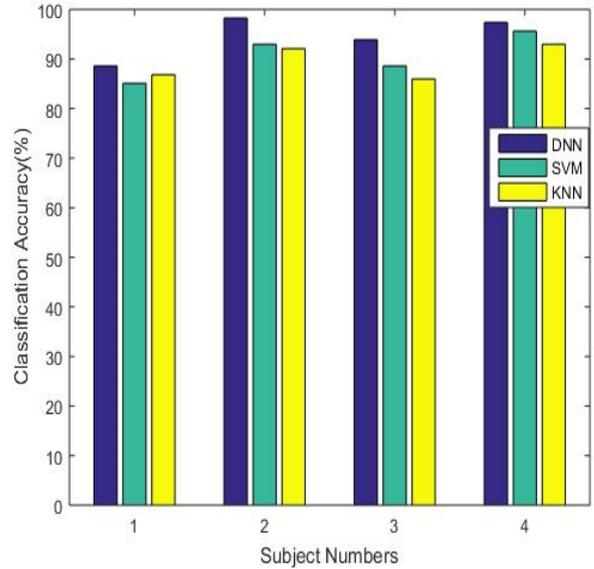

(a) Classification accuracy values for accelerometer input

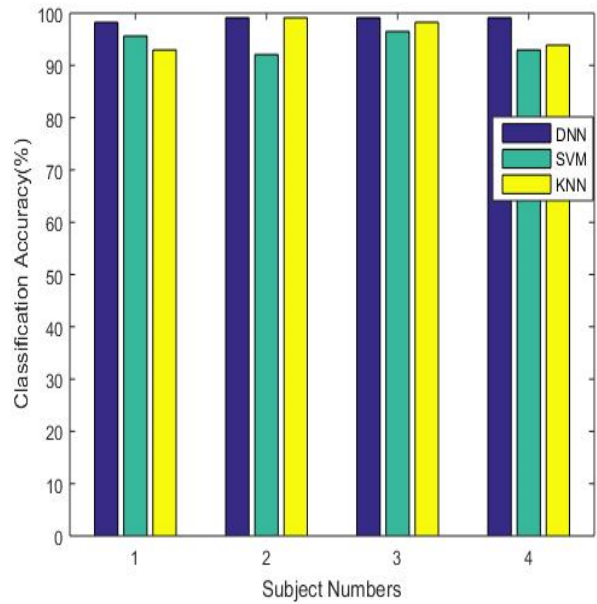

(b) Classification accuracy values for gyroscope input

Fig. 6. Comparison of DNN with conventional algorithms

For each subject, a 3-5% improvement in classification accuracy values is witnessed when comparing DNN and $k$NN. Due to increased training of DNN and optimized parameters used in the design, the classifier distinguishes the hand gestures more accurately. Classification approach using DNN performs better than the conventional algorithms. Due to lack of manual control monitoring of the training process and low depth of the classifiers, SVM and *k*NN fail to remove significant amount of correlation when compared to DNN. Thus, the suitability of the proposed approach is asserted by comparative analysis.

## V. CONCLUSION

In order to produce an efficient algorithm capable of assessing hand activities, a clear distinction in these gestures is of utmost importance. In this work, the hand gestures as input IMU signals from wearable sensors are classified by making use of a novel and improved DNN. Operation of DNN in feed-forward propagation phase as well as back-propagation phase is analyzed and improved. Performance analysis is carried out and presented to depict optimization of the algorithm. Classification using proposed approach is compared to that of conventional algorithms such as SVM and *k*NN. Feature reduction using PCA is carried for each algorithm. Due to manual control of the training process and increased depth, DNN performs better in comparison to SVM and *k*NN in reducing correlation and classifying the hand gestures considered in this paper. A 3-5% improvement in classification accuracy is witnessed on using DNN. Thus, suitability of the algorithm is asserted for IMU signals obtained from wearable sensors.

ACKNOWLEDGMENT

The authors would like to recognize the funding support provided by the Science & Engineering Research Board, a statutory body of the Department of Science & Technology (DST), Government of India, SERB file number ECR/2016/000637.